\documentclass[12pt,fleqn]{iopart}

\usepackage{graphicx}

\begin{document}

\title[Observation of non-classical correlations]{Observation of non-classical correlations in sequential measurements of photon polarization}

\author{Yutaro Suzuki, Masataka Iinuma, and Holger F. Hofmann}

\address{Graduate school of Advanced Sciences of Matter, Hiroshima University, 1-3-1 Kagamiyama, Higashi-Hiroshima 739-8530, Japan
}

\ead{yutaro-s@huhep.org}

\begin{abstract} 
A sequential measurement of two non-commuting quantum observables results in a joint probability distribution for all output combinations that can be explained in terms of an initial joint quasi-probability of the non-commuting observables, modified by the resolution errors and back-action of the initial measurement. Here, we show that the error statistics of a sequential measurement of photon polarization performed at different measurement strengths can be described consistently by an imaginary correlation between the statistics of resolution and back-action. The experimental setup was designed to realize variable strength measurements with well-controlled imaginary correlation between the statistical errors caused by the initial measurement of diagonal polarizations, followed by a precise measurement of the horizontal/vertical polarization. We perform the experimental characterization of an elliptically polarized input state and show that the same complex joint probability distribution is obtained at any measurement strength. 
\end{abstract}


\vspace{2pc}


\maketitle

\section{Introduction}

The relation between non-commuting physical properties remains one of the deepest mysteries of quantum mechanics and is at the heart of many controversies regarding the physics of quantum measurement and quantum information \cite{Oza03,Wat11,Bus13,Dre14,Pus14}. The reason for these controversies is the non-trivial relation between quantum states and the experimentally observable statistics of physical properties. It was already noticed in the early days of quantum mechanics that quantum states can be represented by quasi-probabilities that closely resemble phase space distributions of two conjugate variables which are represented by non-commuting operators in the Hilbert space formalism \cite{Wig32,McCoy32,Kir33,Dir45}. In these quasi-probabilities, the non-classical correlations between physical properties represented by non-commuting Hilbert space operators are represented by the non-positive joint probabilities assigned to the possible combinations of eigenvalues for the two observables. It is clear that such non-positive probabilities do not represent the relative frequencies of measurement outcomes in joint measurements, and this is consistent with the principle that non-commuting observables cannot be measured jointly. However, it is possible to reconstruct joint probabilities by combining a sufficiently large set of uncertainty limited measurements. The first successful application of this principle is the well established method of reconstructing the Wigner function of a single mode light field from measurements of the marginal distributions of field components measured by homodyne detection at different phases \cite{homodyne}. These groundbreaking experiments clearly showed that the negative values of the Wigner function do make a non-trivial statement about the relation between the statistics of non-commuting observables. Unfortunately, the reconstruction of the Wigner function involves data from a large number of non-commuting measurements, making it difficult to identify the relation between experimental statistics and quantum correlations involved in the process. Ideally, non-classical correlations should be observed by measuring two non-commuting observables jointly, for instance in an uncertainty limited sequential measurement. 
For continuous variables of light field modes, this approach can be realized by using a beam splitter to divide the input beam into two parts, followed by measurements of the two non-commuting field components by homodyne detection in the two arms. As a result of the vacuum fluctuations added when the input beam was split, the statistics observed are given by the Husimi- or Q-function, which is equal to the convolution of the Wigner function with a Gaussian representing the uncertainty limited error of a joint quantum measurement of two non-commuting quadrature components of the light field. However, it is not a straightforward matter to extend this theory of sequential measurements to discrete variables, since the assumption that the errors from resolution and back-action are uncorrelated results in measurement outcomes different from the eigenvalues of the discrete observables \cite{Hof00,Hof01}. To realize a direct experimental observation of non-classical correlations in a sequential measurement, it is therefore necessary to properly understand how quantum correlations appear in the joint statistics of resolution errors and back-action errors.  

An interesting solution to the problem of how to measure non-classical correlations is represented by the method of weak measurement, where the back-action of the initial measurement is negligibly small, so that the final outcome corresponds to a precise measurement of the input state and the conditional average of the other observable is obtained by averaging out the statistical errors of the initial weak measurement using data from a very large number of trials \cite{Aha88}. Recently, it has been shown that this method can be used to directly observe the Dirac distribution, which is the quasi-probability that is obtained from products of non-commuting projection operators \cite{Dir45,Joh07,Hof12,Lun12,Boy13,Bam14}. This observation of a quasi-probability is direct in the sense that the individual measurement outcomes all correspond to specific combinations of eigenvalues for the two non-commuting operators. The relation between the experimental data and the reconstructed distribution is established by the measurement errors. In the weak measurement, this is particularly simple because there are no back-action errors, and the statistics of the resolution errors can be obtained from the experimental data. The drawback is that the weak measurement is only an asymptotic limit. Any realistic measurement has a finite strength interaction resulting in a non-vanishing measurement back-action. It is therefore important to understand how the results obtained with weak measurements change as measurement strength increases \cite{Dre10,Hof14a,Zou15,Val16}.

In previous work, we have shown that the errors of a sequential measurement can be explained in terms of the statistics of resolution and back-action errors \cite{Suz12}. In particular, we were able to confirm the violation of Leggett-Garg inequalities by the non-commuting polarization components of photon polarization using the assumption of statistically independent spin-flips. However, a closer analysis of the problem of error statistics shows that the actual correlations between the errors are themselves non-classical and need to be represented by complex error statistics \cite{Kin15}. This result is consistent with the experimental observations of the complex Dirac distribution by weak measurements, since the only way to ensure that the imaginary probabilities of the Dirac distribution are converted into positive relative frequencies for the measurement outcome is the inclusion of an imaginary error probability in the correlations between the resolution errors and the back-action errors. 

In this paper, we present experimental results for a sequential measurement of diagonal and horizontal/vertical photon polarization using an elliptically polarized input state. Since the circular polarizations appear as imaginary correlations in the Dirac distribution of the diagonal and horizontal/vertical polarizations, we need to design a measurement that makes these imaginary correlations appear as real correlations between the initial and final measurement outcomes. We find that this kind of correlation can be realized by introducing a conditional optical phase shift into the variable strength measurement of diagonal polarization. It is then possible to recover the same real and imaginary parts of the Dirac distribution at all measurement strengths, simply by deconvoluting the statistical errors. Specifically, we realize the initial variable strength measurement of diagonal polarization using an interferometric setup, where the input light is first separated into horizontal and vertical polarization components and are then brought to interfere by rotating the polarizations toward the same diagonal axis \cite{Suz12,Iin11}. 
This interference provides the resolution of the initial diagonal polarization measurement in proportion to the amount of the polarization rotations. At the same time, the rotations change the original horizontal and vertical polarization components, resulting in the back-action effect associated with the resolution of the measurement of diagonal polarization. Our setup thus realizes an optimal trade-off between resolution errors represented by the finite visibility of interference between partially distinguishable polarizations and back-action errors represented by the polarization rotations that reduce the distinguishability of polarizations in the two interfering beams. In the present work, we modified the polarization rotation by rotating the linear polarizations in the two beams towards elliptical polarizations circulating in opposite directions. This rotation towards opposite circular polarization components converts circular input polarizations into opposite linear polarizations in the two output ports of the interferometer. As a result, circular input polarizations appear as real correlations between the initial and the final measurements of linear polarization, indicating that the imaginary correlations between the linear polarization components have been successfully converted into an experimentally observable correlation between the initial and the final measurement result. We can then observe the non-classical correlation associated with the non-commutativity of the two polarization components as a statistical correlation between the two measurement outcomes obtained at any combination of measurement resolution and back-action.

The rest of the paper is organized as follows. In section \ref{sec:concepts}, we discuss the relation between the experimental measurement statistics and the quantum statistics of the input state and identify the error statistics of resolution and back-action of the initial variable strength measurement. In section \ref{sec:implement}, we explain how we can implement a measurement with controllable non-classical correlations between resolution and back-action by using a combination of path interference and polarization rotations. In section \ref{sec:exp}, the actual experimental setup is described in detail and its experimentally observed error characteristics are presented. In section \ref{sec:joint}, we present the results obtained for an elliptically polarized input and determine the Dirac distribution by deconvolution of the error statistics observed at each measurement strength. It is shown that the same Dirac distribution is obtained at any measurement strength, confirming the variable strength conversion of imaginary correlations into experimentally observable correlations of the two measurement results. Section \ref{sec:conclusions} summarizes the results and concludes the paper.   

\section{Characterization of errors in a sequential measurement of photon polarization}
\label{sec:concepts}

In the following, we consider a sequential measurement of single photon polarization, where the first measurement distinguishes the diagonal polarizations that correspond to positive (P) and negative (M) superpositions of the horizontal (H) and vertical (V) polarization components of the light field. We can describe the polarizations by self-adjoint operators with eigenvalues of $\pm 1$, specifically
\begin{eqnarray}
\hat{S}_\mathrm{HV} &=& \mid H \rangle\langle H \mid - \mid V \rangle\langle V \mid
\nonumber \\
\hat{S}_\mathrm{PM} &=& \mid H \rangle\langle V \mid + \mid V \rangle\langle H \mid.
\end{eqnarray}
The eigenstates of diagonal polarization $\hat{S}_\mathrm{PM}$ are $\mid P \rangle$ for the eigenvalue of $s_\mathrm{PM}=+1$ and $\mid M \rangle$ for the eigenvalue of $s_\mathrm{PM}=-1$. The initial measurement of diagonal polarization is designed so that we can control the statistics of resolution errors and back-action errors by modifying the measurement interaction. Since there are only two possible measurement outcomes, the errors can be characterized by a single value that determines the precision of the outcome on a scale of zero to one. 

First, we consider the resolution errors. If the measurement outcome is given by $m$, the correct result is obtained if $m=s_\mathrm{PM}$ and a measurement error occurs if $m=-s_\mathrm{PM}$. We can therefore define the resolution $\varepsilon$ as the difference between the probabilities of these two outcomes,
\begin{equation}
\label{eq:defres}
\varepsilon = P(m=s_\mathrm{PM}) - P(m=-s_\mathrm{PM}).
\end{equation}
Experimentally, we can evaluate the resolution $\varepsilon$ by comparing the average result of $m$ with the expectation value $\langle \hat{S}_\mathrm{PM} \rangle$ of the input state,
\begin{equation}
\label{eq:exres}
 \varepsilon = \frac{P_{\mathrm{exp}}(m=+1)-P_{\mathrm{exp}}(m=-1)}{\langle \hat{S}_\mathrm{PM} \rangle_{\mathrm{input}}}.
\end{equation}
In a variable strength setup, the PM resolution $\varepsilon$ increase with the strength of the measurement interaction. As a result, a high resolution can only be obtained in the presence of significant back-action effects. Since we perform a final measurement of horizontal/vertical (HV) polarization, we are especially interested in the effects of the back-action on $\hat{S}_\mathrm{HV}$. A measure similar to the resolution can be defined as the HV transmission $\tau$, which evaluates the fidelity of HV-polarization transmission through the measurement setup in terms of the relation between the input value $s_\mathrm{HV}$ and the final value $f$,
\begin{equation}
\label{eq:deftrans}
\tau = P(f=s_\mathrm{HV}) - P(f=-s_\mathrm{HV}).
\end{equation}
The transmission of HV-polarization can be determined from experimental data by comparing the average value of $f$ with the expectation value $\langle \hat{S}_\mathrm{HV} \rangle$ of the input state,
\begin{equation}
\label{eq:extrans}
 \tau = \frac{P_{\mathrm{exp}}(f=+1)-P_{\mathrm{exp}}(f=-1)}{\langle \hat{S}_\mathrm{HV} \rangle_{\mathrm{input}}}.
\end{equation}
In a variable strength measurement, the transmission fidelity decreases as measurement strength increases. This corresponds to an increase in the rate of back-action induced flips of HV-polarization. 

Resolution $\varepsilon$ and transmission $\tau$ describe the error statistics of measurement and back-action separately. However, a sequential measurement also provides information about the correlations between the non-commuting observables $\langle \hat{S}_\mathrm{PM} \rangle$ and $\langle \hat{S}_\mathrm{HV} \rangle$ in the form of experimentally observed correlations between the measurement outcomes $m$ and $f$. These correlations can be evaluated by taking the product of the two outcomes, $f m = \pm 1$. In principle, one can then define the correlation fidelity $\gamma$ as the difference between the probabilities of obtaining the correct product and the probability of obtaining the opposite value,
\begin{equation}
\label{eq:defcorr}
\gamma = P(f m = s_\mathrm{HV}s_\mathrm{PM}) - P(f m = - s_\mathrm{HV}s_\mathrm{PM}).
\end{equation}
The problem with this definition is that it refers to a product of physical properties that are represented by a pair of non-commuting operators. In the Hilbert space formalism, the product of the two operators is given by
\begin{equation}
\label{eq:xy}
\hat{S}_\mathrm{HV}\hat{S}_\mathrm{PM} = i \hat{S}_\mathrm{RL},
\end{equation} 
where the operator $\hat{S}_\mathrm{RL}$ represents the circular polarization of the photon, with an eigenvalue of $+1$ for right circular polarization (R) and an eigenvalue of $-1$ for left circular polarization (L). The Hilbert space formalism thus suggests that the average product of PM-polarization  $\hat{S}_\mathrm{PM}$ and HV-polarization $\hat{S}_\mathrm{HV}$ is imaginary, with its absolute value of the average given by the circular polarization. In addition, the sign of the imaginary part is related to the operator ordering, which points to a non-statistical origin of this imaginary correlation. 

It is important to recognize that the theoretical formulation of quantum mechanics does not really offer any alternative definitions of a product of two physical properties. The only freedom of choice seems to be the ordering of the operators, which does appear to be arbitrary. However, in the case of a sequential measurement, the ordering can be related to the actual temporal order of the measurements. In Dirac notation, operators act on states to their right, so the operator representing the first measurement should be placed to the right of the operator representing the final measurement. This is indeed the convention used in weak measurements, and this definition of operator ordering defines the ordering in the Dirac distribution obtained in experiments using weak measurements \cite{Lun12,Boy13,Bam14}. In the present case, our goal is to extend these results to variable strength measurements by identifying the quantum correlations between resolution errors and back-action errors. As we know from a general analysis of joint measurements \cite{Kin15}, these correlations correspond to imaginary error probabilities that convert the quantum correlations represented by operator products into experimentally observed correlations between the two measurement outcomes. Specifically, the correlation fidelity $\gamma$ can be determined from experimental data by comparing the average value of the product $f m$ with the expectation value $\langle \hat{S}_\mathrm{HV} \hat{S}_\mathrm{PM}\rangle$ of the input state,
\begin{equation}
\label{eq:imcorr}
 \gamma = \frac{P_{\mathrm{exp}}(f m =+1)-P_{\mathrm{exp}}(f m =-1)}{\langle \hat{S}_\mathrm{HV} \hat{S}_\mathrm{PM} \rangle_{\mathrm{input}}}.
\end{equation}
We can make use of equation (\ref{eq:xy}) to relate the quantum correlation between $\hat{S}_\mathrm{PM}$ and $\hat{S}_\mathrm{HV}$ to the circular polarization of the input, which results in the assignment of an imaginary value to the correlation fidelity $\gamma$, 
\begin{equation}
\label{eq:excorr}
 \nu = i \gamma = \frac{P_{\mathrm{exp}}(f m =+1)-P_{\mathrm{exp}}(f m =-1)}{\langle \hat{S}_\mathrm{RL} \rangle_{\mathrm{input}}}.
\end{equation}
It is therefore possible to determine the imaginary correlation $\gamma =-i \nu$ between resolution errors and back-action errors by comparing the experimentally observed correlation between the outcome $m$ of the variable strength measurement and the outcome $f$ for the HV-polarization after the measurement with the circular polarization in the input. 

Experimentally, the correlation between the two measurement outcomes $m$ and $f$ originates from the back-action effect of the initial measurement of PM-polarization on keeping the HV information of the initial state. If this effect is minimized, the imaginary correlation $\gamma =-i \nu$ will be zero. It is therefore not desirable to realize a minimal back-action of the measurement. Instead, we need to design a measurement with a non-vanishing correlation fidelity, since a non-vanishing value of $\gamma =-i \nu$ is needed for a complete reconstruction of the input state statistics. For an arbitrary input state, we can use the error statistics of the measurement to relate the experimentally observed joint probability $P_{\mathrm{exp}}(m,f)$ to the Dirac distribution that described the quantum state as a complex joint probability $\rho(s_\mathrm{PM},s_\mathrm{HV})$. In general, this relation takes the form of a conditional probability $P_{M}(m,f|s_\mathrm{PM},s_\mathrm{HV})$ \cite{Kin15},
\begin{equation}
\label{eq:general}
 P_{\mathrm{exp}}(m,f) = \sum_{s_\mathrm{PM},s_\mathrm{HV}} P_{M}(m,f|s_\mathrm{PM},s_\mathrm{HV}) \rho(s_\mathrm{PM},s_\mathrm{HV}). 
\end{equation}
Based on the discussion of the experimentally observable errors at the start of this section, we can now express this relation using the resolution $\varepsilon$, the transmission $\tau$, and the imaginary correlation $\gamma =-i \nu$, 
\begin{eqnarray}
\label{eq:flipping}
P_\mathrm{exp}(m,f)  
&=& \frac{1+\varepsilon+\tau - i \nu}{4} \rho(s_\mathrm{PM}=m,s_\mathrm{HV}=f) \nonumber 
\\ &&
 +  \frac{1-\varepsilon+\tau + i \nu}{4} \rho(s_\mathrm{PM}=-m,s_\mathrm{HV}=f) \nonumber 
\\ &&
+  \frac{1+\varepsilon-\tau + i \nu}{4} \rho(s_\mathrm{PM}=m,s_\mathrm{HV}=-f) \nonumber 
\\ &&
+  \frac{1-\varepsilon-\tau - i \nu}{4} \rho(s_\mathrm{PM}=-m,s_\mathrm{HV}=-f).
\end{eqnarray}
The Dirac distribution $\rho(s_\mathrm{PM},s_\mathrm{HV})$ can be reconstructed from the experimental data by inverting this relation,
\begin{eqnarray}
\label{eq:reconstruct}
\rho(s_\mathrm{PM},s_\mathrm{HV})  
 &=& 
 \left(\frac{1}{4}+ \frac{1}{4\varepsilon}+ \frac{1}{4\tau}+ \frac{i}{4\nu}\right) P_\mathrm{exp}(m=s_\mathrm{PM},f=s_\mathrm{HV})  
\nonumber \\ &&
 + \left(\frac{1}{4}-\frac{1}{4\varepsilon}+\frac{1}{4\tau}-\frac{i}{4\nu}\right) P_\mathrm{exp}(m=-s_\mathrm{PM},f=s_\mathrm{HV}) 
\nonumber \\ &&
 + \left(\frac{1}{4}+\frac{1}{4\varepsilon}-\frac{1}{4\tau}-\frac{i}{4\nu}\right) P_\mathrm{exp}(m=s_\mathrm{PM},f=-s_\mathrm{HV}) 
\nonumber \\ &&
 + \left(\frac{1}{4}-\frac{1}{4\varepsilon}-\frac{1}{4\tau}+\frac{i}{4\nu}\right) P_\mathrm{exp}(m=-s_\mathrm{PM},f=-s_\mathrm{HV}).
\end{eqnarray} 
In principle, a complete reconstruction is possible whenever the coefficients $\varepsilon$, $\tau$ and $\nu$ are all non-zero. In practice, small values in any of these sensitivities of the measurement will result in an amplification of statistical errors, so there will be practical limitations on the reconstruction of the Dirac distribution in the extreme cases of both very weak and very strong measurements. Nevertheless it is possible to reconstruct the complete quantum statistics of the input state for the full range of measurement strengths between the extreme limits from the correlated outcomes of a sequential measurement of the two non-commuting properties $\hat{S}_\mathrm{PM}$ and $\hat{S}_\mathrm{HV}$, where the initial measurement introduces resolution and back-action that change the initial complex probabilities of the Dirac distribution into experimentally observable correlations between the initial and final measurement outcomes $m$ and $f$.   

\section{Implementation of sequential measurement sensitive to the imaginary correlation} 
\label{sec:implement}

The sequence of a measurement of PM-polarization followed by a measurement of HV-polarization is implemented by using a variable strength measurement of PM-polarization, where the two possible outcomes correspond to the two output ports of an optical interferometer. Specifically, we exploit the fact that the diagonal polarizations correspond to interferences between the HV-polarization components in the input to achieve active control over the measurement strength by transferring polarization coherence to path coherence \cite{Iin11}. After separating the H and the V polarization components spatially, path interference is made possible by rotating the polarizations in the two paths towards each other. In previous work \cite{Suz12,Iin11,Iin16}, we used this method to optimize the uncertainty trade-off between resolution and back-action, which is accomplished by rotating the polarizations towards a common diagonal polarization. Specifically, the rotation of each HV component results in a partial randomization of the HV information of the initial polarization state, which is the back-action effect of the PM-measurement. By rotating the polarizations towards a common diagonal polarization, this back-action can be converted into an interferometric measurement of the initial PM-polarization, where the visibility of the interference determines the resolution of the measurement. 

In the present experiment, we modify this method by rotating the polarizations towards elliptical polarizations with the same major axis of polarization but opposite circular polarizations. Since the circular polarizations in the two paths are opposite, they do not contribute to the path interference at the output that distinguishes P-polarization from M-polarization. Instead, the effect of the twist towards opposite circular polarizations can only be seen when the effect on the HV-polarization in the output is taken into account. Specifically, the interference at the final beam splitter introduces a correlation between the final HV-polarization and the output port that represents the result of the PM-measurement. Thus the twist of the polarization rotation changes the correlations between the outcomes of the sequential measurements, modifying the sensitivity of the setup to non-classical correlations between the PM-polarization measured by the interferometric setup and the HV-polarization measured in the output. 

As mentioned above, the first step in the realization of the PM-measurement is the separation of H-polarization and V-polarization at a polarizing beam splitter. It is then possible to control the back-action coherently by rotating the polarization in the two branches away from their original polarization by a rotation angle of $2 \theta$. In addition to the magnitude of this rotation angle, we can also control the direction of the rotation. Experimentally, we achieve this control by sandwiching a rotatable half-wave plate (HWP) between two $\lambda / 8$-wave plates (OWPs) with opposite alignments parallel to the HV-polarization axes. Ideally, the OWPs change the direction of polarization rotation induced by the HWP towards elliptical polarizations exactly halfway between the PM-polarizations and the RL-polarizations. In general, we can describe the direction of polarization rotation by an additional angle $\phi$ corresponding to the rotation angle around the HV-axis of the Bloch sphere. We can then describe the coherent transmission of the measurement setup for the two output paths of the interferometer. Using the operators $\hat{M}_{\mathrm{P}}$ and $\hat{M}_{\mathrm{M}}$ to represent the output ports for the measurement results P and M, respectively, the coherent transmission effects on the input state can be written as
\begin{eqnarray}
\label{eq:Minter}
\hat{M}_{\mathrm{P}} \mid H \rangle &=& \frac{1}{\sqrt{2}}\left(\cos(2 \theta) \mid H \rangle + \exp(i \phi) \sin(2 \theta) \mid V \rangle\right)
\nonumber 
\\
\hat{M}_{\mathrm{P}} \mid V \rangle &=& \frac{1}{\sqrt{2}}\left( \exp(i \phi) \sin(2 \theta) \mid H \rangle + \cos(2 \theta) \mid V \rangle\right)
\nonumber 
\\
\hat{M}_{\mathrm{M}} \mid H \rangle &=& \frac{1}{\sqrt{2}}\left(\cos(2 \theta) \mid H \rangle - \exp(i \phi) \sin(2 \theta) \mid V \rangle\right)
\nonumber 
\\
\hat{M}_{\mathrm{M}} \mid V \rangle &=& \frac{1}{\sqrt{2}}\left(-\exp(i \phi) \sin(2 \theta) \mid H \rangle +  \cos(2 \theta) \mid V \rangle\right).
\end{eqnarray}
These four coherent transmission functions completely define the measurement operators. In particular, they describe an operation that leaves the PM-polarization unchanged. It is therefore possible to represent equation~(\ref{eq:Minter}) in more compact form using $\hat{S}_{\mathrm{PM}}$ and the identity $\hat{I}$, 
\begin{eqnarray}
\label{eq:Mops}
 \hat{M}_{\mathrm{P}} &=& \frac{1}{\sqrt{2}}\left(\cos(2\theta)\hat{I} + e^{i\phi} \sin(2\theta) \hat{S}_\mathrm{PM} \right) \nonumber \\
 \hat{M}_{\mathrm{M}} &=& \frac{1}{\sqrt{2}}\left(\cos(2\theta)\hat{I} - e^{i\phi} \sin(2\theta) \hat{S}_\mathrm{PM}\right), 
\end{eqnarray}
where $\theta$ is the rotation angle of the HWP and $\phi$ is the phase shift induced by the OWP. By using OWPs, the angle $\phi$ is ideally fixed at $\phi=\pi/4$, which should provide an optimal compromise between measurement resolution and sensitivity to imaginary correlations.  

The settings $\theta$ and $\phi$ allow us to coherently control the experimentally observable statistics of the measurement. Since the change in HV-polarization only depends on the HWP angle $\theta$, the most direct control is achieved for the transmission fidelity $\tau$ with
\begin{equation}
\label{eq:tau}
\tau_{\mathrm{ideal}} = \cos(4 \theta).  
\end{equation}
The measurement resolution $\varepsilon$ depends on the both angles of the HWP $\theta$ and the $\phi$ of the polarization rotation with
\begin{equation}
\label{eq:epsilon}
\varepsilon_{\mathrm{ideal}} = \cos(\phi) \sin(4 \theta).  
\end{equation}
Note that this ideal value is only obtained for optimal visibility of the interferometer. Likewise, interference is necessary to observe the correlation fidelity $\nu=i \gamma$. The value obtained for perfect interference is
\begin{equation}
\label{eq:nu}
\nu_{\mathrm{ideal}} = - \sin(\phi) \sin(4 \theta).  
\end{equation}
Thus the parameter $\phi$ distributes the coherent effects of the measurement between resolution error and imaginary error correlation, thereby controlling the trade-off between sensitivity to PM-polarization and sensitivity to non-classical correlations in the input.

\section{Experimental setup}
\label{sec:exp}

\begin{figure}[th]
\centering
 \begin{center}
\includegraphics[width=110mm]{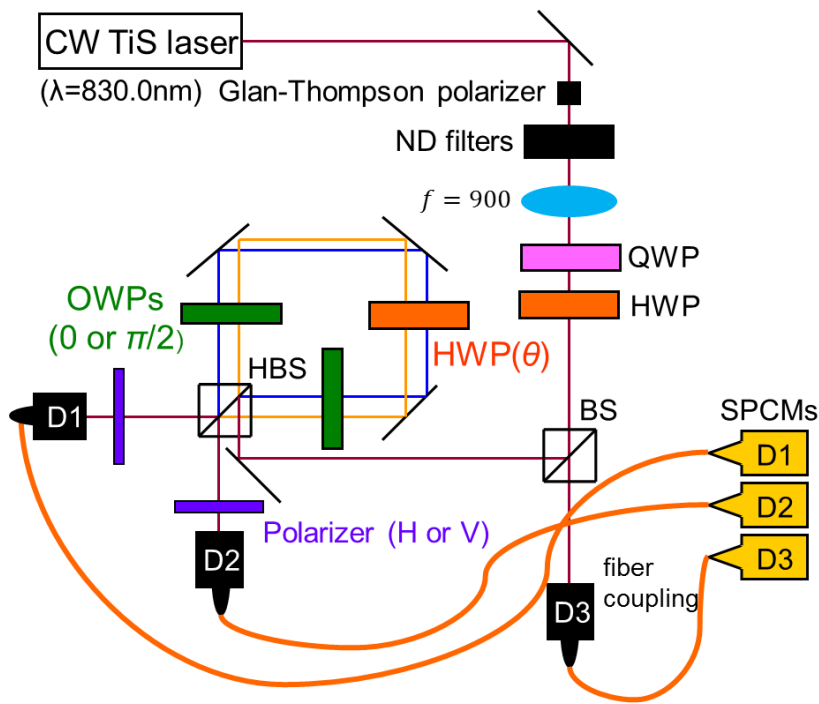}
 \end{center}
 \caption{Experimental setup for the sequential measurement of PM and HV polarization. A Sagnac interferometer with a hybrid-coated beam splitter operating as polarizing beam splitter in the input and as polarization insensitive 50:50 beam splitter in the output is used to separate and interfere the horizontal and vertical polarization components. The measurement strength is adjusted by the rotation angle of the half-wave plate inside the interferometer, with the two $\lambda /8$-wave plates changing the polarization rotation to induce sensitivity to the non-classical correlations between PM-polarization and HV-polarization in the input.}
 \label{fig:setup}
\end{figure}

The experimental setup for the sequential measurement of photon polarization is shown in figure~\ref{fig:setup}. It is similar to the setup used previously to investigate the trade-off between resolution errors and back-action errors \cite{Suz12,Iin11}, with the important difference that the polarization rotation introduces additional correlations between measurement errors and back-action. The variable strength measurement of PM-polarization is realized by the central Sagnac interferometer. At the input port of the interferometer, the photon path is split into H and V component paths by a hybrid-coated beam splitter (HBS), which acts as a polarizing beam splitter for the input beam. Inside the Sagnac interferometer, the counter-propagating beams pass through two $\lambda / 8$-wave plates(OWPs) and a half-wave plate(HWP), such that the HWP is sandwiched between the two OWPs. The two OWPs placed on both sides of the HWP are aligned along the axes of HV-polarization, with the fast axis of one OWP aligned along H and the fast axis of the other OWP aligned along V. The HWP can be rotated, and its rotation angle $\theta$ determines the strength of the back-action by transforming H-polarization into V-polarization and vice versa. Since the two optical paths go through the same HWP from opposite directions, a single HWP set at an angle of $\theta$ is sufficient to rotate the polarizations in both of the two counter-propagating beams. By sandwiching the HWP between the OWPs, the total operation performed by the sequence of three optical elements corresponds to a polarization rotation towards elliptical polarizations along the same diagonal P direction, but with opposite circular polarization components for the H and the V inputs.  

Input photons are generated using attenuated light from a CW titanium-sapphire laser (wavelength 830.0nm, power 900mW). The light is first passed through a Glan-Thompson polarizer to select only H-polarized photons. Neutral density (ND) filters reduce the intensity to permit the detection of individual photons. Arbitrary initial polarizations can be prepared by a combination of quarter-wave plate (QWP) and HWP upstream of the interferometer. For the detection of HV-polarization, polarization filters were inserted into the output beams. Output photons were detected by single photon counting modules (SPCM-AQR-14). Typical count rates were 150kHz. To compensate possible fluctuations in the rate of input photons, the input beam was divided by an additional beam splitter upstream on the interferometer and the count rate of the beam split off from the input was monitored. A lens was inserted downstream of the ND filters to adjust the beam profile and to optimize the path interference at the beam splitter part of the HBS. 

\begin{figure}[ht]
\centering
\includegraphics[width=70mm]{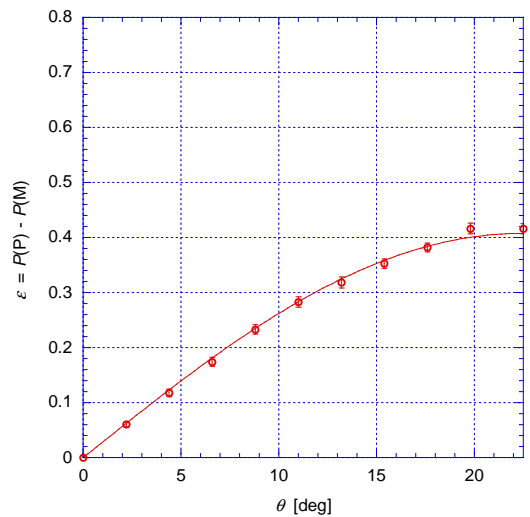}
 \caption{Experimental evaluation of measurement resolution $\varepsilon$ using P-polarized input photons.The measurement strength $\theta$ is given by the degrees of rotation for the half-wave plate in the interferometer.}
 \label{fig:resolve}
\end{figure}

The statistics of measurement errors can be evaluated experimentally based on the definition of resolution $\varepsilon$, transmission $\tau$, and imaginary correlation fidelity $\gamma$ in equation~(\ref{eq:defres}),~(\ref{eq:deftrans}),~and (\ref{eq:defcorr}) respectively. For the resolution of the PM-measurement, we used a P-polarized input and evaluated the probabilities of the measurement outcomes P and M to obtain 
\begin{equation}
\varepsilon = P(\mbox{P}|\mbox{P}) - P(\mbox{M}|\mbox{P}).
\end{equation}
Note that this result distinguishes only the output ports of the interferometer, and is independent of the polarization of the output photons. The result for different measurement strengths is shown in figure~\ref{fig:resolve}. As expected, the dependence on measurement strength $\theta$ can be described by a sine function with its maximal value $0.416 \pm 0.009$ at 22.5 degrees or $\theta=\pi/8$. We can fit the data with
\begin{equation}
\varepsilon(\theta) = V_\varepsilon \sin(4 \theta),
\end{equation}
where $V_\varepsilon$ corresponding to the maximal resolution observed at $\theta=\pi/8$ is $0.408 \pm 0.004$. 

\begin{figure}[ht]
\centering
\includegraphics[width=70mm]{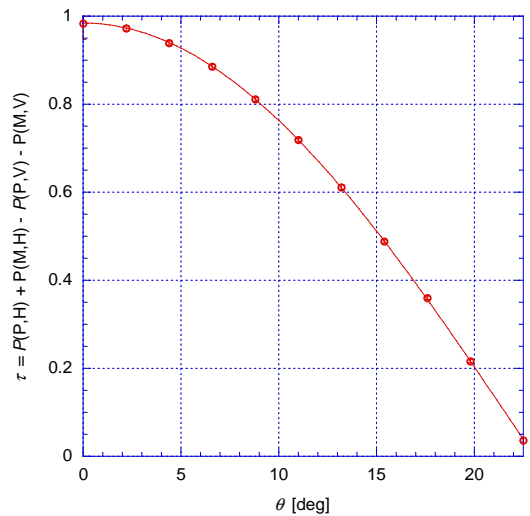}
 \caption{Experimental evaluation of transmission $\tau$ as a function of measurement strength $\theta$ using H-polarized input photons.}
 \label{fig:transmit}
\end{figure}

A similar characterization can be performed for the transmission $\tau$. In this case, we used an H-polarized input and the HV-polarization was measured in both output ports. The total transmission fidelity is then given by
\begin{equation}
\tau =P(\mbox{P},\mbox{H}|\mbox{H}) + P(\mbox{M},\mbox{H}|\mbox{H})- P(\mbox{P},\mbox{V}|\mbox{H}) - P(\mbox{M},\mbox{V}|\mbox{H}).
\end{equation}
The experimental results obtained at different measurement strengths is shown in figure~\ref{fig:transmit}. Since the back-action error is directly induced by the rotation of the HWP, the dependence on measurement strength $\theta$ is close to the theoretical ideal given by equation~(\ref{eq:tau}). At $\theta=0$, the transmission is $\tau(0)=0.98 \pm 0.04$. However, we also find a non-vanishing difference between the HV-polarizations in the output at a maximal measurement strength of $\theta=\pi/8$. Although the origin of this deviation from the expected result is not entirely clear, it may be helpful to consider the possibility that the control in liner polarization is not precise, since $\tau=0$ requires a rotation of the initial HV-polarization into an elliptical polarization oriented exactly along the diagonal between H and V. The experimental value is $\tau(\pi/8)=0.036 \pm 0.004$, which corresponds to the theoretical value at an HWP angle of $\theta=0.977 \times (\pi/8)$ or 21.98 degrees. The precise value of $\tau$ may therefore be difficult to control at high measurement strengths. 

\begin{figure}[ht]
\centering
\includegraphics[width=70mm]{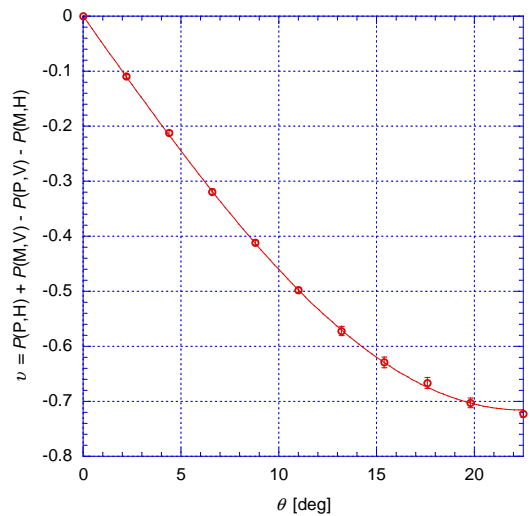}
 \caption{Experimental evaluation of correlation fidelity $\nu=i \gamma$ as a function of measurement strength $\theta$ using R-polarized input photons. The negative sign indicates that a positive imaginary correlation appears as an anti-correlation between the sequential outcomes in the experimental measurement statistics.}
 \label{fig:correlate}
\end{figure}

Finally, we characterize the most important element of the error statistics, the correlation fidelity $\nu=i \gamma$. As shown in equation~(\ref{eq:defcorr}) and~(\ref{eq:excorr}), the correlation fidelity is observed by taking the expectation value of the product of the two measurement outcomes in the sequential measurement. The input correlation depends on the circular polarization and has a value of $+i$ for R polarization. We can therefore determine the correlation $\nu$ directly by evaluating
\begin{equation}
\nu =P(\mbox{P},\mbox{H}|\mbox{R}) + P(\mbox{M},\mbox{V}|\mbox{R})- P(\mbox{P},\mbox{V}|\mbox{R}) - P(\mbox{M},\mbox{H}|\mbox{R}).
\end{equation}
Importantly, both the output ports and the final HV-polarization are completely random. The only characteristic feature of the probability distribution of output results is the correlation between the output ports of the PM-measurement and the HV-polarization in the output, which originates from the non-classical correlation between the polarizations associated with the circular polarization of the input. $\nu$ thus quantifies the conversion of unobservable imaginary correlations into experimentally observed correlations between the two measurement outcomes. The experimental results are shown in figure~\ref{fig:correlate}. Similar to the resolution $\varepsilon$, the dependence on measurement strength can be described by a sine function. The values of $\nu$ are negative, because the direction of polarization rotations in the interferometer results in an anti-correlation between the measurement outcomes for the positive imaginary input correlations given by an R-polarized input. This maximal negative value is $0.723 \pm 0.007$ at $\theta=\pi/8$. We can fit the data with 
\begin{equation}
\nu(\theta) = - V_\nu \sin(4 \theta),
\end{equation}
where $V_\nu$ indicates the maximal negative value of $\nu$ that we expect to achieve in our setup. According to the fit of the data shown in figure~\ref{fig:correlate}, the value of $V_\nu$ is $0.716 \pm 0.003$.

Experimentally, both the resolution $\varepsilon$ and the correlation $\nu$ depend on the visibility of interference at the output beam splitter of the interferometer. This means that the total visibility is given by $\sqrt{V_\varepsilon^2+V_\nu^2}=0.824\pm0.003$, which corresponds well with the visibility of 0.82 observed in a direct characterization of the interferometer used in our setup. For $\phi=\pi/4$, we would theoretically expect equal values for $\varepsilon$ and $\nu$. The difference between the values suggests that the rotation angle $\phi$ around the HV-axis actually had a value of about 60.3 degrees. The experimental characterization of the setup therefore suggests that the birefringent phase shift in the Sagnac interferometer was actually significantly larger than the 45 degrees expected from the use of $\lambda/8$-wave plates. 

By characterizing the measurement statistics experimentally, we can obtain the correct reconstruction procedure for an arbitrary input state at any measurement strength. Most importantly, we can observe the imaginary correlation between PM-polarization and HV-polarization as a real correlation between the initial and the final measurement outcome in the sequential measurement, where the magnitude of the observed correlation is a well-defined function of measurement strength. It is therefore possible to verify that the correlation between the two measurement results originates from the correlation between measurement errors and back-action in the initial measurement of PM-polarization.

\section{Experimental statistics of non-classical correlations in sequential measurements}
\label{sec:joint}

In the previous section, it was shown that our experimental setup is sensitive to both the linear and the circular polarization of the input, where the circular polarization determines the correlations between the measurement outcomes of the two measurements of linear polarization. It is therefore possible to reconstruct the complete quantum statistics of an arbitrary input state from the joint probabilities of experimental outcomes obtained in a  sequential measurement of two non-commuting properties. We demonstrate this possibility by choosing a right circulating elliptically polarized input state with the major axis of the ellipse oriented halfway between H-polarization and P-polarization. 

\begin{figure}[ht]
\centering
\includegraphics[width=100mm]{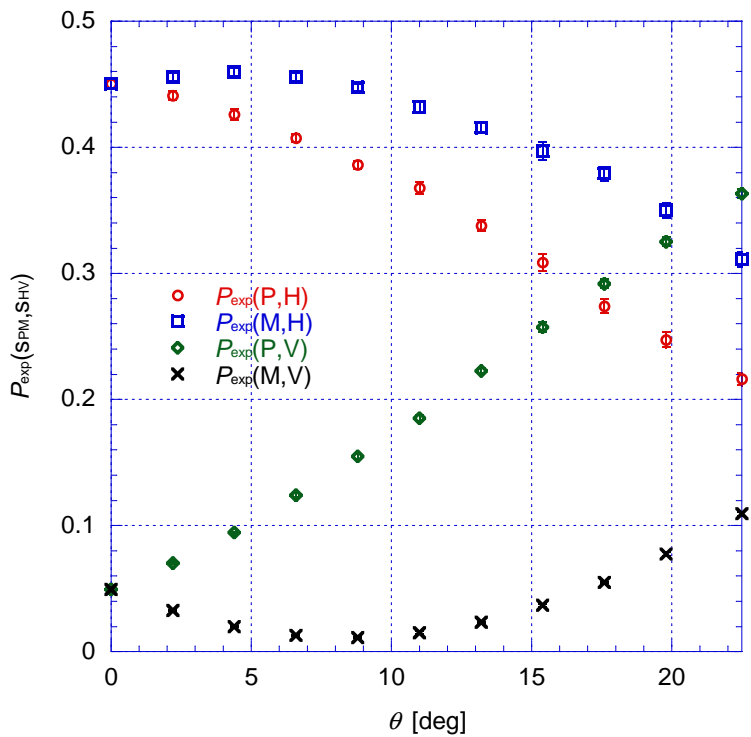}
 \caption{Experimental joint probabilities for an elliptically polarized input state at different measurement strengths $\theta$}
 \label{fig:expprob}
\end{figure}

Figure~\ref{fig:expprob} shows the results for the joint probabilities $P_{\mathrm{exp}}(s_\mathrm{PM},s_\mathrm{HV})$ of the four experimental outcomes observed in the experiment for different measurement strengths $\theta$. The characteristics of the input state are clearly visible at all measurement strengths. At low measurement strength, the final HV-measurement is nearly error free and shows the expected preference for H-polarization, with a probability of 90\% for H-polarization in the limit of $\theta=0$. As measurement strength increases, we expect that the P-polarized outcomes become more likely than the M-polarized outcomes, reflecting the preference for P-polarization in the input state. However, we find that the maximal probability in the low to intermediate range of measurement strengths is obtained for the combination (M,H), which has a higher probability than the combination of the two most likely polarizations (P,H) at all measurement strengths. As measurement strength increases from $\theta=0$ to $\theta= 5$ degrees, the probabilities of (M,H) and of (P,V) increase, while the probabilities of (P,H) and (M,V) drop. Thus the measurement statistics show a strong preference for the two outcomes with $s_{\mathrm{PM}}s_{\mathrm{HV}}=-1$ over the two outcomes with $s_{\mathrm{PM}}s_{\mathrm{HV}}=+1$. This trend continues even at higher measurement strengths. In fact, it is especially obvious in the statistics at $\theta > 17$ degrees, where (M,H) and (P,V) are the two most likely outcomes with probabilities that both exceed the probability of (P,H). Thus, the statistics at $\theta > 17$ degrees is dominated by the correlation product of the measurement outcomes, and depends much less on the individual values observed separately in the initial and the final measurement. 

The change in measurement strength corresponds to a change in sensitivity from HV-polarization towards PM-polarization. However, full resolution of PM-polarization is never obtained due to the trade-off with correlation sensitivity, as shown in figures~\ref{fig:resolve} and~\ref{fig:correlate}. As a result, the region of high measurement strength is dominated by the observation of correlations between the two measurement outcomes, where the negative average value of $s_{\mathrm{PM}}s_{\mathrm{HV}}$ originates from the right circular polarization in the input state. As discussed in section~\ref{sec:concepts}, it is possible to express the quantum state in terms of the Dirac distribution of PM-polarization and HV-polarization, where the circular polarization appears as an imaginary correlation between the two linear polarization components. Since the experimental results already have the form of a joint probability of HV-polarization and PM-polarization, we only need to invert the matrix representing the statistical errors of the measurement process to obtain the Dirac distribution from the experimental data, as shown in equation~(\ref{eq:reconstruct}). The coefficients that describe the measurement errors are determined for each measurement strength $\theta$ using the data shown in figures~\ref{fig:resolve},~\ref{fig:transmit} and~\ref{fig:correlate}. Thus the reconstruction serves as a test of the assumption that the separate characterization of measurement errors in section~\ref{sec:exp} is also valid for an arbitrary input state. 

\begin{figure}[ht]
\centering
\includegraphics[width=100mm]{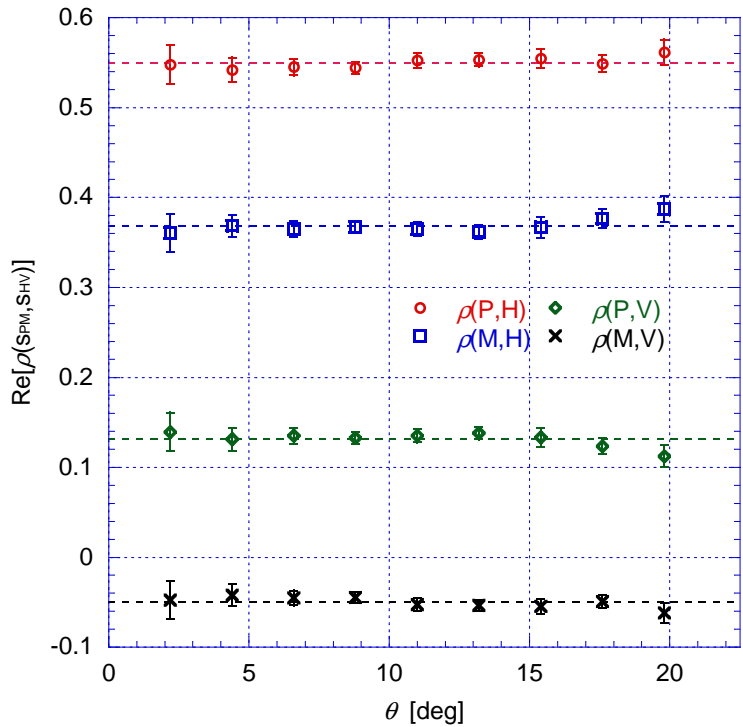}
 \caption{Real part of the Dirac distribution obtained from the experimental joint probabilities at different measurement strengths. As indicated by the dashed lines, the results obtained at different measurement strengths are all explained by the same input state statistics.}
\label{fig:rhoRe}
\end{figure}

The result for the real part of the Dirac distribution is shown in figure~\ref{fig:rhoRe}. The error bars shown in the figure have been obtained by combining statistical errors in the experimental data with statistical errors in the estimated values of $\varepsilon$ and of $\tau$. The latter errors increase in both the weak and the strong limit, since the reconstruction of PM and HV statistics becomes very sensitive to the precise values of the very small resolution $\varepsilon$ and transmission $\tau$ in these limits. As indicated by the dashed lines, the results obtained at different measurement strengths reproduce nearly the same initial quantum statistics, confirming that the measurement strength dependence of the data in figure~\ref{fig:expprob} is sufficiently explained by the variation of the error statistics characterized in section~\ref{sec:exp}. The real part of the Dirac distribution shows the linear polarization of the state, as given by the expectation values of $\hat{S}_\mathrm{PM}$ and $\hat{S}_\mathrm{HV}$. The average results indicated by the dashed lines in figure~\ref{fig:rhoRe} correspond to expectation values of $\langle \hat{S}_\mathrm{PM} \rangle = 0.36$ and $\langle \hat{S}_\mathrm{HV} \rangle = 0.84$, which shows that the major axis of the elliptically polarized input was somewhat closer to H-polarization than to P-polarization. Note that the algebra of equation~(\ref{eq:reconstruct}) ensures that the real parts of $\rho(\mathrm{P,H})$ and $\rho(\mathrm{M,V})$ and the real parts of $\rho(\mathrm{M,H})$ and $\rho(\mathrm{P,V})$ each have a sum of 0.5, so that the correlations between PM-polarization and HV-polarization in the real part of the Dirac distribution have no state-specific meaning. The correlations between the initial outcome of the PM-measurement and the final outcome of the HV-measurement only contribute to the evaluation of the imaginary correlations between the two non-commuting properties in the Dirac distribution. 

 \begin{figure}[ht]
\centering
\includegraphics[width=100mm]{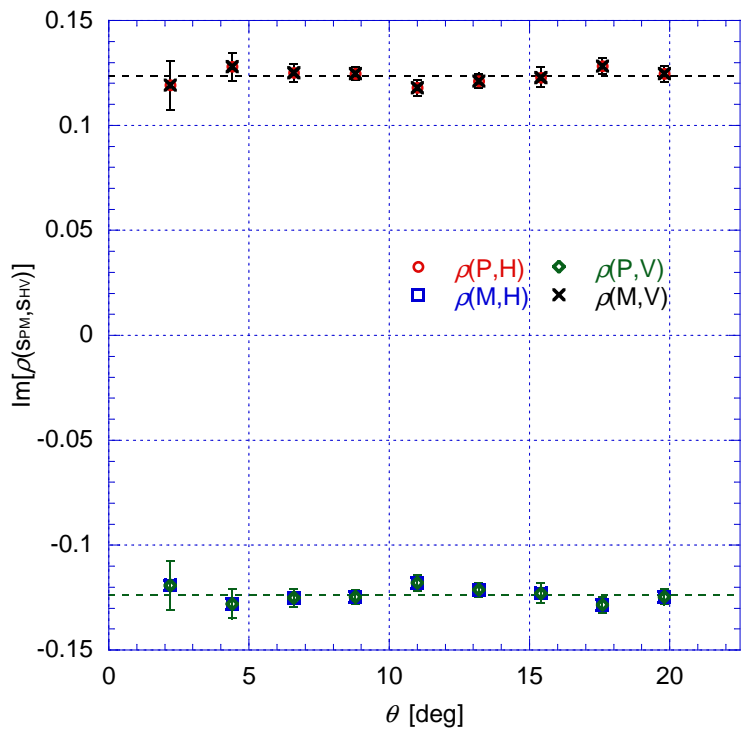}
 \caption{Imaginary part of the Dirac distribution obtained from the experimental joint probabilities at different measurement strengths. Since marginal probabilities must be real, $\mbox{Im}[\rho(\mathrm{P,H})]=-\mbox{Im}[\rho(\mathrm{M,H})]=-\mbox{Im}[\rho(\mathrm{P,V})]=\mbox{Im}[\rho(\mathrm{M,V})]$.}
\label{fig:rhoIm}
\end{figure}

Figure~\ref{fig:rhoIm} shows the results for the imaginary parts of the Dirac distribution that are reconstructed using the experimentally observed value of the correlation fidelity $\nu$ at different measurement strengths $\theta$. The results confirm that the same input statistics is observed at all measurement strengths, and that the correlation observed between the PM-outcomes and the HV-outcomes can be traced back to the circular polarization of the input photons once the measurement strength dependence of the errors is taken into account. Specifically, the imaginary statistics reconstructed from the experimental data corresponds to an expectation value of $\langle \hat{S}_\mathrm{HV} \hat{S}_\mathrm{PM}\rangle = i 0.50$, which is the characteristic imaginary correlation of the right circulating elliptically polarized input state.  

To understand the significance of the present experiment, it is important to remember that the imaginary value of the Dirac distribution in figure~\ref{fig:rhoIm} has been obtained from the experimentally observed correlations between the two measurement outcomes shown in figure~\ref{fig:correlate}. The reconstruction of an input state with $\langle \hat{S}_\mathrm{RL} \rangle=0.50$ therefore confirms that the experimentally observed correlations between the outcomes of the PM-measurement and the outcomes of the HV-measurements originate from the non-classical correlation expressed by the operator product $\hat{S}_\mathrm{HV} \hat{S}_\mathrm{PM}=i \hat{S}_\mathrm{RL}$. A sequential measurement performed at any measurement strength provides the complete statistics of the input state in the form of a joint probability of experimental results for the two non-commuting observables, where the statistical correlations between the measurement results originate from the non-classical correlations described by ordered operator products. Sequential measurements thus provide a particularly direct method of quantum state tomography, where the raw data obtained for the output probabilities is closely related to the quasi-probability description provided by the Dirac distribution.

\section{{Conclusion}}
\label{sec:conclusions}

We have realized a sequential measurement of two non-commuting polarization components of a single photon. It is shown that the joint probabilities obtained at any measurement strength provide a complete map of the input state statistics, as given by the Dirac distribution of the two non-commuting observables. In particular, we can show that the correlations between the initial and the final measurement result can be traced back to the imaginary correlations described by the product of the non-commuting operators by taking into account the imaginary error correlations that describe the quantum dynamics of the initial measurement.

The main experimental achievement is the implementation of a variable strength measurement with a non-vanishing dynamical correlation between the resolution errors and the back-action of the measurement. In optics, this is possible by implementing the measurement using quantum interferences between the coherent implementation of back-action in the paths of an interferometer that effectively separates the eigenstates of the observable whose disturbance by the back-action will be detected in the output. We can then realize a trade-off between the resolution of the initial measurement and the conversion of imaginary correlations into experimentally observable real correlations by the back-action. 

It is important to note that the dynamics of the measurement interaction is closely related to the non-classical aspects of the statistics represented by the non-commutativity of the operators \cite{Hof11,Hof14b,Hof16}. This is clearly not just a technical problem, but relates directly to the constraints that quantum mechanics imposes on our intuitive notion of reality. The complex and non-positive statistics of the Dirac distribution cannot be reconciled with any joint reality of the two non-commuting physical properties, even though the joint statistics can be identified as the origin of the randomness in all actual measurement results. The ability to vary the distribution of errors in a joint measurement allows us to demonstrate that the non-classical joint statistics of non-commuting observables represented by the Dirac distribution provides an objective and measurement independent description of the relation between the physical properties observed in the actual experimental outcomes. Seemingly paradoxical statistics described by imaginary and negative parts of the quasi-probabilities are possible and indeed necessary, because the quantum dynamics of the measurement converts them into real contributions to actual probabilities observed as relative frequencies of the joint outcomes \cite{Hof15}. The fact that the outcomes are obtained in sequence is useful because it allows us to identify the origin of this conversion between non-classical correlations and experimental results as part of the interaction dynamics of the initial measurement. This identification with the dynamics also provides an explanation for the operator ordering in the non-classical correlation, since it identifies the measurement sequence with the sequence in which the operators are multiplied with each other. 

In conclusion, the present work helps to illustrate how quantum statistics determine the outcomes of measurements that are simultaneously sensitive to different non-commuting observables. The results obtained about the relation between the imaginary expectation values of operator products and the correlations between resolution errors and back-action in quantum measurements provide important evidence for the role of quantum dynamics in the definition of measurement uncertainties. The analysis presented in this paper can therefore serve as a starting point for a more comprehensive exploration of the possibilities and limitations of control in quantum systems.  

\section*{Acknowledgments}
This work was supported by Grant-in-Aid for JSPS Fellows number 2605259 and JSPS KAKENHI Grant No. 24540428.

\end{document}